# Vortical flow structures induced by red blood cells in capillaries

François Yaya,[1, 2,] Johannes Römer,[3] Achim Guckenberger,[3] Thomas John,[1] Stephan Gekle,[3] Thomas Podgorski,[2] and Christian Wagner[4, 5 a)]

[1)]*Saarland University, Experimental Physics, Campus building E2 6, Saarbrücken, Germany*
[2)]*Laboratoire Interdisciplinaire de Physique, 140 rue de la Physique, Saint Martin d'Hères, France*
[3)]*Biofluid Simulation and Modeling, Fachbereich Physik, Universität Bayreuth, Bayreuth, Germany*
[4)]*Saarland University, Experimental Physics, Campus building E2 6, Saarbrücken, Germany*
[5)]*Physics and Materials Science Research Unit, University of Luxembourg, 162a avenue de la Faïencerie, Luxembourg*

*a) Corresponding author: c.wagner@mx.uni-saarland.de*

ABSTRACT

**Objective:** Knowledge about the flow field of the plasma around the red blood cells in capillary flow is important for a physical understanding of blood flow and the transport of micro- and nanoparticles and molecules in the flowing plasma. We conduct an experimental study on the flow field around red blood cells in capillary flow that are complemented by simulations of vortical flow between red blood cells.

**Methods:** Red blood cells were injected in a 10x12 µm rectangular microchannel at a low hematocrit and the flow field around a single or two cells have been characterized thanks to a highspeed camera and by tracking 250 nm nanoparticles in flow behaving as tracers.

**Results:** While the flow field around a steady "croissant" shape is found to be relatively similar to that of a rigid sphere, the flow field around a "slipper" shape exhibits a small vortex at the rear of the red blood cell. Even more pronounced are vortex-like structures observed in the central region between two neighboring croissants.

**Conclusions:** The rotation frequency of the vortices is to a good approximation, inversely proportional to the distance between the cells. Our experimental data are confirmed and complemented by numerical simulations.

Abbreviations: BI, boundary integral; BIM, boundary integral method; BSA, Bovine Sulfate Albumine; PBS, phosphate buffer saline; PDMS, polydimethylsiloxane; RBCs, red blood cells;

I. INTRODUCTION

Besides its physiological relevance, blood flow in microcapillaries is a prime example of a biological fluid structure interaction problem between the elastic red blood cells (RBCs) on the one hand and the hydrodynamic flow of plasma on the other [1,2,3]. Over recent years, quite some attention has been paid to the dynamics of RBCs in cylindrical or rectangular channels known to be the most common configuration in model microfluidic flows. Depending on external parameters such as flow speed, channel size and plasma viscosity two main shapes have emerged from these experimental [4,5,6,7,8,9] as well as numerical [9,10,11,12,13,14,15,16,17,18] works. The first, called slipper, is an elongated, non-axisymmetric shape in which the RBC tends to flow at a steady position slightly away from the channel center. The second one, almost axi-symmetric in cylindrical capillaries and with two planes of symmetry in rectangular channels, is termed "parachute" in cylindrical tubes or "croissant" in rectangular cross-sections and flows in the channel center. In addition, some works observed clusters of two or more RBCs formed and held together by hydrodynamic interactions [19,20,21,22,23,24,25,26]. While there is good agreement between numerical simulations and experiments regarding the actual shape of RBCs, less attention has been paid to resolve the flow field of the surrounding plasma experimentally. While in the absence of RBCs, this would be a simple parabolic Poiseuille profile, the presence of RBCs strongly disturbs the flow [27]. Understanding the actual flow pattern is however essential, e.g., for the distribution of nanometric drug delivery agents or dissolved substances in microcapillary blood flow [26,28,29,30,31,32,33,34] as well as to understand the nature of interactions between cells. For a simple sphere in a cylindrical channel the flow pattern can be computed analytically, but nevertheless leads to surprisingly intricate dynamics of the solute [35]. While there is quite some literature on the flow field in microchannels around rigid or only slightly deformable objects such as microspheres or droplets [36,37,38,39], experimental data are rare for the complex flows arising due to the above described croissant and slipper motions of red blood cells. In RBC clusters, computer simulations have indicated the presence of vortex-like structures between neighboring RBCs [19,20,22,26]. Here, we present a direct experimental observation of (i) the flow around isolated RBCs and (ii) the vortical flow between clustering RBCs in a rectangular microchannel using particle tracking. We find that the rotation frequency of the vortices scales inversely proportional to the RBC distance. Our particle tracking measurements are in good agreement with corresponding boundary-integral simulations.

## II MATERIAL AND METHODS

### A. Experiments

To mimic blood capillaries, we use rectangular straight channels of 12 µm width, 10 µm depth and about 40 mm in length as microfluidic chips. The camera is positioned such that the long (12 µm) side is viewed in the images. Channels are made from Polydimethylsiloxane (PDMS) [40]. To avoid RBCs adhering onto the walls, the channels are flushed with a buffer solution containing BSA at 1%. The flow of the suspension is observed in a microscope with an oil-immersion objective (Nikon CFI Plan Fluor 60x, NA = 1.25). The field of observation is at 1 cm behind the pressure inlet. We use a high speed camera (HiSpec Fastec 2G) to record image sequences at frame rates of 9000 fps. We investigate the flow at cell speeds at the physiologically relevant parameter range and beyond from 1 mm/s up to 10 mm/s. The various flow and cell speeds are achieved with a pressure controller (ELVEFLOW OB1+) at pressure drops ranging from 100 mbar to 1000 mbar, respectively [9].

To visualize the flow field around the moving RBC, we add nanoparticles with surface coating of polyethylene glycol (PEG) and 250 nm in diameter (MICROMOD) in an aqueous solution containing 0.1% of Bovine Serum Albumine (BSA). Blood is drawn from healthy donors after giving an informed consent in compliance with the ethical requirements of the Saarland University, Saarbrücken, Germany (Ärztekammer des Saarlandes, approval number 24/12). We use a suspension of washed RBCs in Phosphate Buffered Saline (PBS) at 0.5% hematocrit with nanoparticles as tracers. Samples are regularly mixed to prevent the sedimentation of RBCs. The shape and speed of individual cells depend on the applied pressure drop and remains stationary within the time of observation in our experiments. We observe two main steady cell shapes in our rectangular channel geometries, croissants at rather low flow speeds and slippers at higher flow speeds, see Fig. 1. We determine the speed of the RBCs from the recorded image sequence over a distance of 100 μm. The trajectories of the tracers are determined with a self-made MATLAB program in the co-moving frame of the individual RBC.

Figures 1(a) and 1(b) (Multimedia view) show exemplary trajectories as overlayed images of snapshots of a flowing RBC for two different velocities. The motion of tracers between two RBC are studied in a similar manner. During the observation timespan, the distance $d_{RBC}$ between two consecutive RBC never changes more than 10% in our experiments, i.e. both cells have comparable velocities.

### B. 3D Simulations

In brief, our 3D boundary integral (BI) simulations solve the Stokes equations for the fluid inside and outside the red blood cell [9,41,42] which is justified by the small Reynolds number of around 0.1. A specific advantage of BI simulations is that the full instantaneous flow field at any time can be computed knowing only the parabolic background flow and the forces on the cylinder wall and the RBC membrane [41,42]. A time integration is not necessary to compute the flow field. In our model, the membrane forces are computed following the models by Skalak and Helfrich for shear, area and bending elasticity, respectively [2,43,44]. The no-slip boundary condition as well as continuity of stresses is imposed at the membrane. Periodic boundary conditions along the channel are used with a computation window length of 42 µm. This length is sufficient to recover a nearly undisturbed Poiseuille flow far away from the cell. For the velocity field computation, a snapshot of the simulation is selected and the shape of the RBC at this time is frozen. Then, the velocity field is computed on a regular grid in the central plane and transformed to the frame of reference of the moving RBC using its center-of-mass velocity. As the slipper exhibits periodic oscillation through a series of slightly varying shapes, an arbitrary shape out of these is selected for computing the streamlines following the procedure described above. In an ideal situation, the mirror symmetry of the system around the central plane would forbid the existence of out-of-plane currents. Due to rounding and discretization errors, nevertheless small spurious out-of-plane currents may arise which however we set to zero here.

The simulations are performed in the same channel geometry as the experiments. Figures 1(c) and 1(d) show the converged 3D RBC shapes and corresponding stream lines in the channel middle plane for a croissant and a slipper, respectively. To compute the stream lines between two cells in a cluster, we select a converged croissant shape from the simulation with a single RBC at the corresponding speed. This shape is then copied, and both cells are placed into the channel. After a short timespan of typically 70 ms, the streamlines are computed starting at a vertical line between the two cells. The above procedure had to be chosen because in the experiments many different distances between two cells for a given flow speed can be observed which means that the clusters are not yet at their converged distance. The reference frame is the center-of-mass velocity of the left RBC. Streamlines contain only directional information but no absolute velocities which are needed to compute the rotation period of particles trapped in flow vortices. To obtain this information, we place virtual tracer particles (again starting from a line of seed points) and integrate their trajectory in time given the flow field extracted from BIM simulations (which is assumed to be stationary). This procedure allows us to obtain not only the particle trajectories, but also temporal information such as rotation periods. In addition, Fig. 5 shows particle trajectories integrated backward in time in order to demonstrate the possibility of particle escape from the central vortices in non-perfectly symmetric situations.

III. RESULTS

## A. Flow field in the vicinity of a single RBC

Fig. 1 shows the experimental and numerical results of our flow field measurements around the two characteristic cell shapes. Far away from the cell, the velocity profile in the rectangular channel is almost parabolic [45]. The speed of the RBCs is approximately the mean velocity of the free liquid flow in the channel because the volume flux is conserved. The cell speed defines for the co-moving frame and is always lower than the fluid velocity in the middle of the channel far away from the cell.

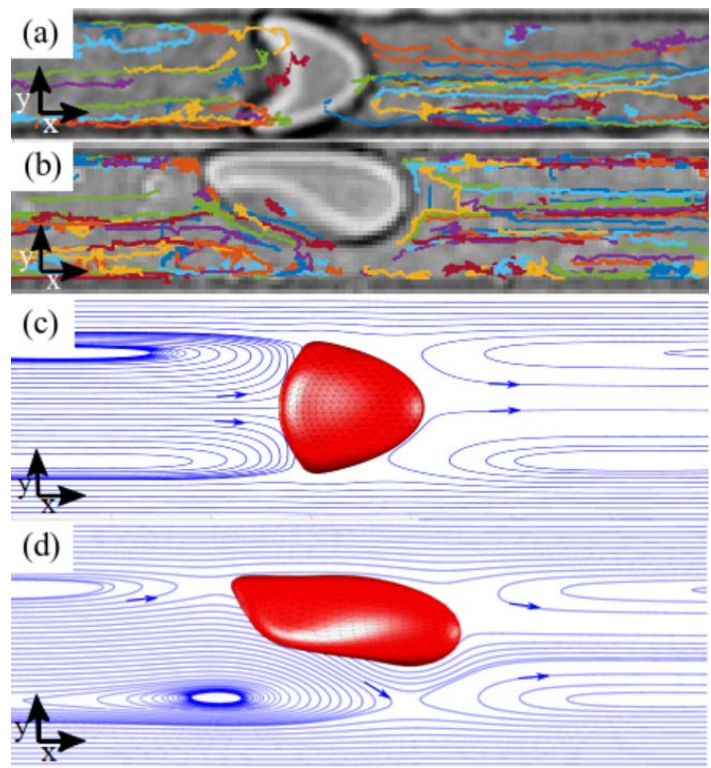

**Fig. 1** Snaphots of RBCs in (a) croissant and (b) slipper shape, together with an overlay of experimental recorded trajectories of tracers. (c) and (d) represent the streamlines and cell shapes from the numerical simulations in the central plane of the channel. The speed in (a) and (c) is $v_{RBC}$ = 2.83mm/s and in (b) and (d) $v_{RBC}$ 6.50mm/s. Particles trajectories and streamlines are shown in the co-moving frame of the RBCs. The fluid flows from left to right. Experimental videos from (a), (b) can be found in supplementary material (Multimedia view)

At low flow speeds, the observed cell shape is in the symmetric croissant shape Fig. 1(a) and 1(c) [9] and tracers in the middle of the channel in front of the cell move away from the cell, while behind the cell they move towards it as shown in Fig. 2(a). In the co-moving frame the flow velocity is zero at the cell surface and there is a stagnation point on the cell surface on the center line. Closer to the walls, this flow direction is reversed. The combination of those motions lead to a strongly elongated half ellipse for the trajectory of a single tracer. These streamline are comparable to what was observed experimentally for flowing droplets in microchannel [38].

From the experimental particle tracking velocimetry (PTV) of the tracers, we can also deduce the relative velocity in the co-moving frame $\vec{v_r}$. The component in flow direction $v_{r,x}$ is shown as function of time in Figure 2(b).

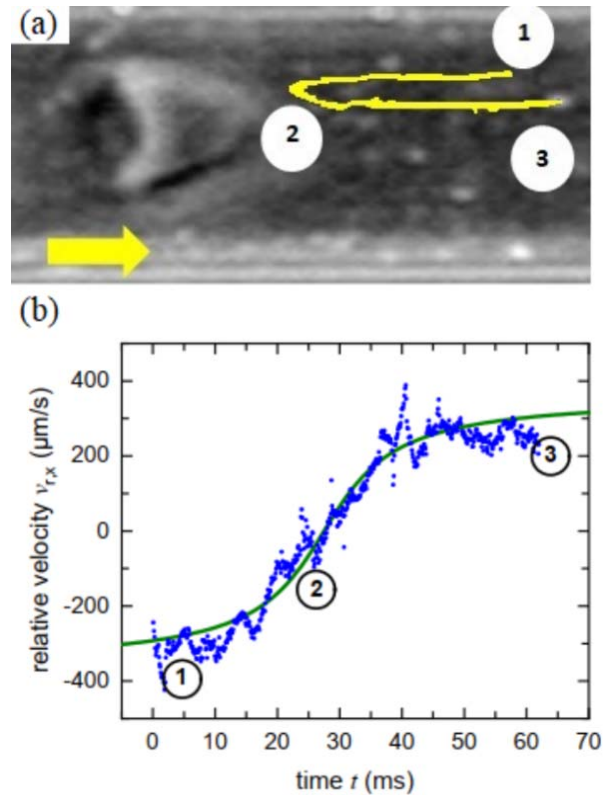

**Fig. 2** Snapshots of RBCs in (a) croissant and (b) slipper shape, together with an overlay of experimental recorded trajectories of tracers. (c) and (d) represent the stream lines and cell shapes from the numerical simulations in the central plane to the channel. The speed in (a) and (c) is $v_{RBC}$ = 2.83mm/s and in (b) and (d) $v_{RBC}$ = 6.50mm/s. Particles trajectories and streamlines are shown in the co-moving frame of the RBCs. The fluid flows from left to right. Experimental videos from (a), (b) can be found in supplementary material (Multimedia view)

The velocity can be fitted with a heuristic sigmoidal function. Obviously, tracers decelerate when approaching the cell and accelerate to the faster fluid motion in the middle of the channel when they move away again from the cell. We should mention that with our microscopic setup in the experiment we image the full sample height and therefore the observed position of tracers is always their projection in the x-y plane. Therefore, we also observe trajectories that approach and depart from the cell in the y = 0 plane (Fig. 1(a)). This is especially the case when the tracer's motion is in the orthogonal plane with respect to the plane of projection. At higher flow speeds, the cells attain a slipper shape [9]. In this asymmetric configuration, a vortical flow can be observed behind the cell (see Fig. 1(b)) even though the flow remains laminar. The slipper shape and the existence of the vortices are confirmed by our numerical simulations (Fig. 1(d)). However, a fully closed vortex flow in the middle plane of the channel as predicted from the numerical simulations, could not be seen in the experimental situation where small irregularities are enough to displace tracer particles out of the vortex zone in the channel. This indicates that in realistic situations it will be unlikely that particles should get trapped in such a vortex. In both cases, for croissants as well as for slippers, the presence of RBCs reduce the spatial variations in fluid velocities over the channel width to an almost plug like flow and therefore the spreading of suspended tracers in flow direction over the channel is also reduced. This becomes even more pronounced for clusters of two RBCs as we will see in the following.

### B. Flow field in the vicinity of a cluster of two RBCs

At low hematocrit levels, the mean distance between consecutive RBCs in the channel is typically very large compared to the cell size and the channel width. If two cells come close to each other, however, they can form a stable cluster due to their hydrodynamic interaction [19,20,21,22,23,24,25,26]. A significant amount of the liquid between the cells seems to be 'encapsulated' and does not mix anymore with the liquid outside of the cluster. Similar to the tracer trajectories of single cells, we observe hairpin loops and vortices for tracers moving in the plane of observation and straight lines for tracers moving orthogonal to it (Fig. 3(a) (Multimedia view)), as consequence of our optical projection. Our numerical simulations confirm that the streamlines correspond to toroidal vortex, just as smoke rings, with the axis of symmetry along the channel x-direction, see Fig. 3(b). In the co-moving frame, the channel borders move with the speed of the RBC-cluster in the − x direction. This relative border motion drives the rotation of the liquid between the cells, similar to the case of a lid driven cavity flow [46] and those types of vortices have been observed for the case of flowing droplets in microfluidic devices [39]. Accordingly, the liquid in the inner part of the channel moves in the +x direction with a speed that is somewhat higher than the speed of the cells in the laboratory frame.

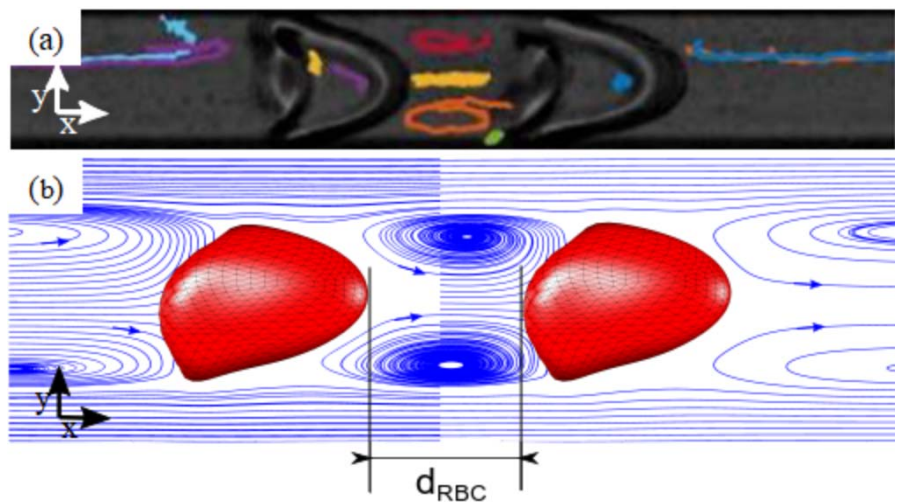

**Fig. 3** (a) Snapshot of a cluster of two RBCs in croissant shape with a distance of $d_{RBC}$ = 5.2 µm together with an overlay of tracer trajectories. The liquid between the cells seems to be encapsulated in a vortex and to rotate as a torus with the axis of symmetry in x-direction. (b) Numerical simulation of a cluster. The streamlines indicate how tracers will be transported from the middle along a helical trajectories (see Fig.6). An experimental video from (a) can be found in supplementary material (Multimedia view)

At steady state, the distance $d_{RBC}$ between two RBCs in a cluster is constant and depends on the flow velocity. However, in the experiments many distortions can disturb the equilibrium distance positions and different distances can be observed for the same flow condition. To mimic this situation in the simulations, we place two cells with the approximated shape at desired distance in the channel and determine the flow field after a short time span of typically 70 ms. Here, we analyze only the flow field of clusters of two croissant shaped cells. To characterize the torus movement of the tracers more quantitatively we consider the period T for one cycle. This means that the vortex forms a deformed torus in 3D and that the flow becomes faster with increasing distance from the central axis of revolution. Fig. 4 shows the period T as a function of the distance between the two cells, $d_{RBC}$. The period increases almost linearly with the distance, both for the simulated and for the experimental results. In other words, in first approximation, the tracers move along the circumference C of ellipses. Assuming a constant tracer speed $v_t$ along a particular ellipse circumference the period $T(d_{RBC})$ is given by the ratio of C and $v_t$. Within concentric ellipses and with decreasing sizes, the velocity $v_t$ decreases as well. The extension of the largest ellipse is given by an almost constant semi-minor axis a in order of a quarter of the channel width and a major axis $d_{RBC}$ of the interior distance between two cells. Using a truncated series expansion for the circumference C of the largest ellipse (details in the Supporting Information), the estimated period T(d$_{RBC}$) is given by:

$$T_{(RBC)} = \frac{C}{v_t} = \frac{\pi}{v_t}\left(a + \frac{d_{RBC}}{2}\right)\left(1 + \frac{\theta^2}{4}\right) \quad (1)$$

With: $\theta = (a - d_{RBC}/2)(a + d_{RBC}/2)$

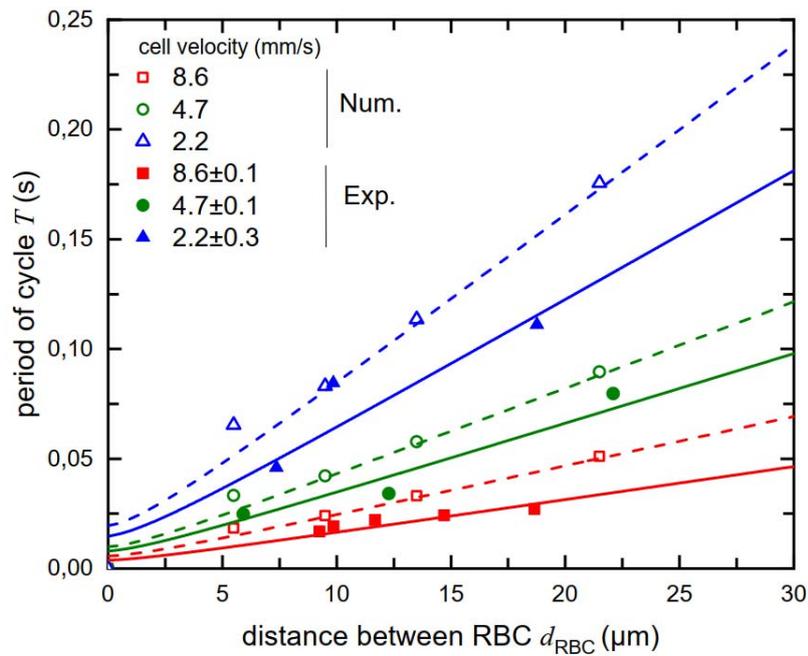

**Fig. 4** Period of cycles for a toroidal tracer motion between two RBCs as a function of the distance between the cells. Experimental data are represented by solid symbols and lines and numerical results by open symbols and dashed lines respectively. The fit represents a motion on an axial stretched torus with ellipses of revolution. It is based on equation (1) with $\pi/v_t$ as a fitting parameter and the minor axis is fixed to be $a = 3\mu m$ (see supplementary material).

This relationship is confirmed by both experiments and simulations. In fact, both show that tracer velocities $v_t$ in the comoving frame happen to be lower than the velocity of a cell $v_{RBC}$ in the laboratory frame. However, there are some quantitative differences which we attribute to the fact that the cells at various distances are not in their equilibrium situation. We use this dependency to extract the fluid and tracer speed $v_t$ at the surface of the toroidal vortex, $v_t$ (see supplementary material). We show in Fig. 4 the extracted periods as function of the distance between two RBC flowing at different speeds. As expected, an offset can be observed due to the fact that the semi-minor axis a is considered constant. We found in simulations that proper vortices were rarely formed for $d_{RBC} > 40\mu m$. In this case, a tracer escapes the vortex before doing a full revolution. For distances shorter than 5 μm, no vortex could be observed in experiments. This is in agreement with simulations [19] who define the "vortex existence" boundary to be 1.4 $R_0$ with $R_0$ being the effective radius of a RBC.

### C. Particle escaping pathways around RBC

Given the vortex-like structures between two clustering RBCs, one may expect that these vortices should be able to actually capture small particles and trap them between the cells similar to earlier observations for microparticles [26]. In figure 5 we illustrate such events, i.e., the pathways by which particles can enter into the vortices from the main flow. For this, a line of tracer particles has been seeded into the central region between the two RBCs and their trajectories integrated backward in time. Thus, the colored lines illustrate the entry of tracer particles into the vortices.

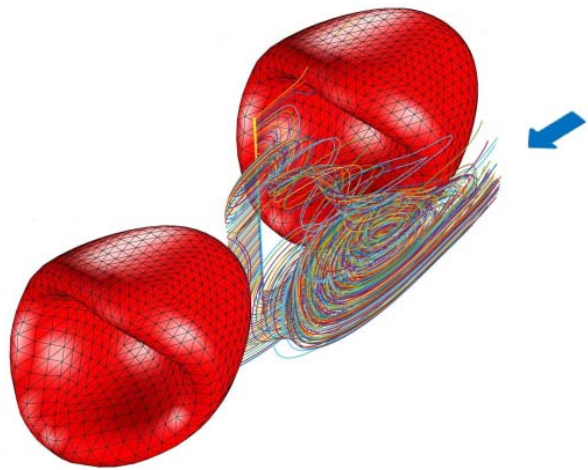

**Fig. 5** 3D view showing the provenance (blue arrow) of the tracers and their trajectories for a cell velocity of 2.2 mm/s.

As can be seen in the streamlines in figure 3, the vortices between two flowing RBCs typically are not fully closed. One can therefore expect that tracer particles should, depending on their starting position, also be able to escape from these vortices back into the main flow. Indeed, we do observe such trajectories as illustrated in figure 6. Note that, in particular for the experiments the period T is not affected by the projective view. For a fixed distance between the cells $d_{RBC}$, we find that the period T remains constant, independent of the amplitude in x-direction. This result is supported by numerical simulations (see Fig. 6). Here, a particle first performs a spiralling motion between the two RBCs but eventually escapes from the region between the two RBCs. Due to the short observation time, we were unfortunately not able to observe such trajectories the experiment.

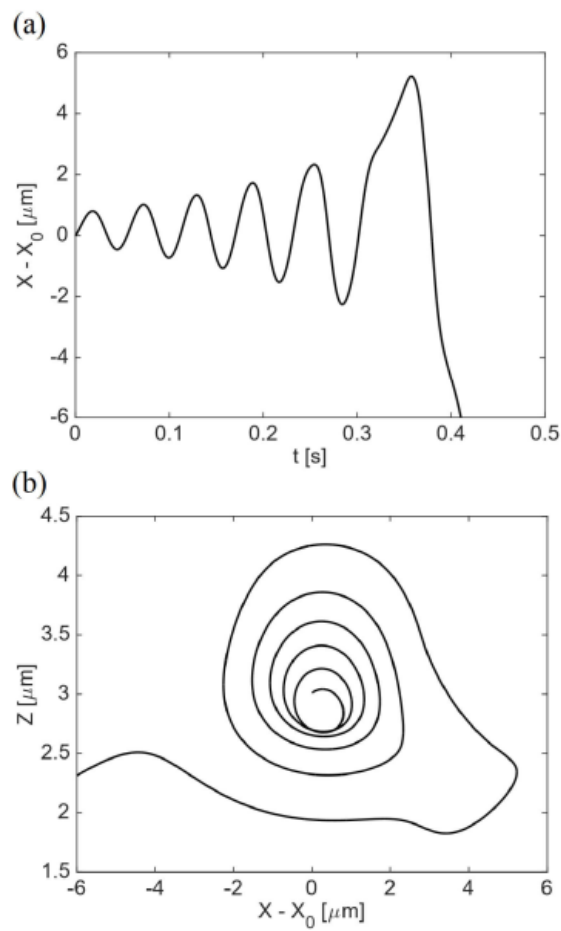

**Fig. 6** Numerically calculated trajectories of a tracer escaping from flow vortices between two RBCs with velocity 2.2 mm/s where $X_0$ is the center between the two

RBCs. (a) Temporal forward-backward motion along the direction of the flow. (b) Toroidal trajectory in the x-y plane, starting from the middle and going outwards.

## IV. DISCUSSION

Using PTV experiments we measured the flow field around red blood cells flowing in small microchannels. For isolated RBCs, we found a significant difference in the flow field depending on whether the red blood cell is in the croissant or in the slipper state. At the rear of a slipper a small but characteristic vortex-like structure is observed. If two cells flow in close vicinity to each other in the croissant shape, another toroidal vortex-like structure appears between the cells as predicted by earlier numerical simulations [19,20,22,26]. However, while there in earlier work numerical simulations of flowing RBCs have been compared with experiments only with respect of the shape of the RBCs we here also compare the surrounding flow field. We showed that the rotation period in these vortices follows a quasi-linear law as a function of the distance between the red blood cells. Our experimental results are in good agreement with numerical boundary-integral simulations. The simulations, in addition, could illustrate qualitatively the trapping and escape of tracer particles from the vortex structures between the cells, which in principle could contribute to the transport and mixing properties of potential nanoparticle drug delivery agents. However, even in our artificial in-vitro setup the flow seems to be not stable enough that we could reproduce these events experimentally and trapping of particles due to the flow field of single or diluted RBCs might be rarely the case in-vivo.

PERSPECTIVES

We give experimental evidence of the presence of vortices around different shapes of red blood cells in capillary flow and we characterize their velocity. The experimental findings are confirmed and extended by numerical simulations. This provides the basis for a quantitative description of capillary blood flow and drug delivery by micro- or nanoparticles.

ACKNOWLEDGMENTS

F.Y., T.J. and C.W. acknowledge funding from the French German University (DFH / UFA). C.W. acknowledges funding from the DFG FOR 2688 - Wa1336/12. F. Y. and T. P. acknowledge support from CNES and LabEx Tec21. S.G. acknowledges funding from the DFG FOR 2688, project B3 (Projektnummer 417989940 - GE2214/2-1). We gratefully acknowledge computing time provided by the SuperMUC system of the Leibniz Rechenzentrum, Garching, and by the Bavarian Polymer Institute.

# Supplementary Material: Vortical flow structures induced by red blood cells in capillaries

François Yaya,[1, 2,] Johannes Römer,[3] Achim Guckenberger,[3] Thomas John,[1] Stephan Gekle, [3] Thomas Podgorski, [2] and Christian Wagner[4, 5]

[1)]Saarland University, Experimental Physics, Campus building E2 6, Saarbrücken, Germany
[2)]Laboratoire Interdisciplinaire de Physique, 140 rue de la Physique, Saint Martin d'Hères, France
[3)]Biofluid Simulation and Modeling, Fachbereich Physik, Universität Bayreuth, Bayreuth, Germany
[4)]Saarland University, Experimental Physics, Campus building E2 6, Saarbrücken, Germany
[5)]Physics and Materials Science Research Unit, University of Luxembourg, 162a avenue de la Faïencerie, Luxembourg

## I.   FITTED TRACER SPEEDS TO ELLIPSES

The observed traces of the nanoparticles between two RBC could be approximated to ellipses in 3D, and the speed was almost constant along the path. However, due to the limited observation time of a RBC pair in the field of view, mostly only a part of the ellipse was detected in the experiments. The projection of those ellipses from 3D to the 2D observation remains as (deformed) elliptical shape and in particular, the period of an orbit is not affected from the projection. All possible ellipses are encapsulated by the channel walls and the surfaces from the the two RBCs. Smaller ellipses inside the biggest ellipse were also observed and exhibited approximately the same period for a cycle. We suggest a rough estimation of the period of the circulations as function of the distance between the cells **d$RBC$** to fit our measurements.

We assume a constant tracer speed **v$_t$**, a fixed semi-minor axis **a** = 3 **µ**m as a quarter of the channel width and the semi-major axis is **b** = **d$_{RBC}$** /2, the half of the distance between the cells. The circumference **C** of those ellipses can be calculated from an expansion series:

$$C = \pi(a + b) \left( 1 + \frac{\theta^2}{4} + \frac{\theta^4}{64} + \cdots \right) \quad (1)$$

$$\theta = (a - b) / (a + b) \quad (2)$$

The $\theta^4$ term is negligible. Based on a constant tracer speed $v_t$ the period for a cycle is given in Eq. (1) in the
main text. For $v_{RBC}$ = 2.2; 4.7; 8.6 *mm/s*, the fitted $v_t$ were respectively: 0.283; 0.556; 0.976 mm/s in simulations and 0.364; 0.664; 1.415 mm/s in experiments. The most extreme case for a particles trace is, when the particle comes very close to the wall. In those case, the speed of the tracer is zero in laboratory frame and $-v_{RBC}$ in co-moving frame respectively. Of course the fitted speeds were always slower as those upper limit.

## II. FLOW FIELD VIEWED FROM DIFFERENT AXES

The parachute shapes of RBC in circular capillaries and quadratic channels are rotational symmetric. Our channels are rectangular in experiments as well in simulations, therefore the shape is slightly squeezed in the smaller direction and the shape is called croissant. For completeness, we therefore show in Fig.1 the streamlines for a single croissant, a slipper and a cluster of two croissants in the **z-x-plane**.

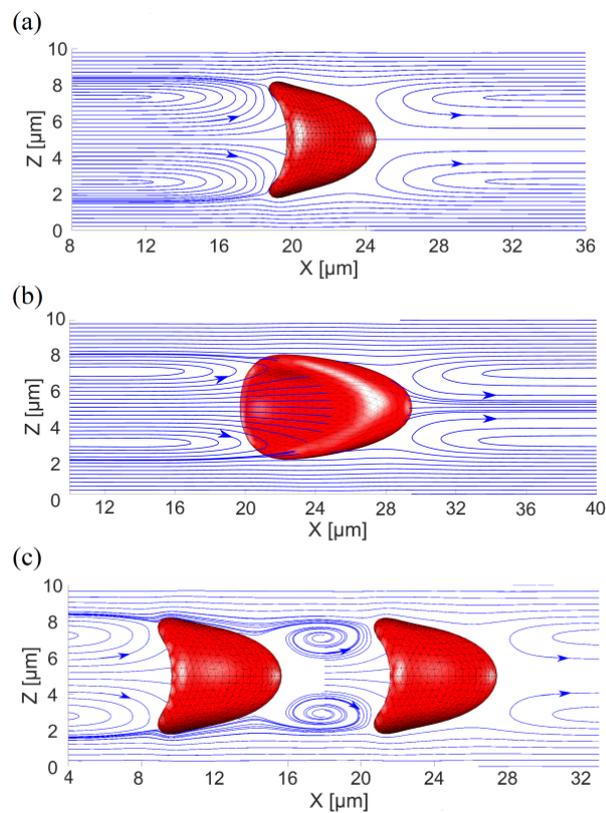

**Fig. 7** (a) Equivalent of Fig. 1(a) and (b) equivalent of Fig.1(a) in the main text, but shown are the stream lines in the z-x plane. (c) Equivalent of Fig. 3(b) in the main text, but shown are the stream lines the z-x plane.

### III. VIDEOS

The videos can be accessed on the Journal "Microciruclation" Website.

croissant.avi A video of a croissant flowing at $v_{RBC}$ = 2.83 mm/s and tracer particles in its vicinity. The duration of the movie in real time is 92 ms.

slipper.avi A video of a slipper flowing at $v_{RBC}$ = 6.50 mm/s and tracer particles in its vicinity. The duration of the movie in real time is 41 ms.

cluster.avi A video of a cluster of two croissants flowing at $vRBC$ = 3.64 mm/s and tracer particles showing the 3D structure of vortices in between. The duration of the movie in real time is 68 ms.

FIGURE REFERENCES

Fig 1. Snaphots of RBCs in (a) croissant and (b) slipper shape, together with an overlay of experimental recorded trajectories of tracers. (c) and (d) represent the streamlines and cell shapes from the numerical simulations in the central plane of the channel. The speed in (a) and (c) is $v_{RBC}$ = 2.83mm/s and in (b) and (d) $v_{RBC}$ 6.50mm/s. Particles trajectories and streamlines are shown in the co-moving frame of the RBCs. The fluid flows from left to right. Experimental videos from (a), (b) can be found in supplementary material (Multimedia view)

Fig. 2 Snapshots of RBCs in (a) croissant and (b) slipper shape, together with an overlay of experimental recorded trajectories of tracers. (c) and (d) represent the stream lines and cell shapes from the numerical simulations in the central plane to the channel. The speed in (a) and (c) is vRBC = 2.83mm/s and in (b) and (d) vRBC = 6.50mm/s. Particles trajectories and streamlines are shown in the co-moving frame of the RBCs. The fluid flows from left to right. Experimental videos from (a), (b) can be found in supplementary material (Multimedia view)

Fig. 3 (a) Snapshot of a cluster of two RBCs in croissant shape with a distance of $d_{RBC}$ = 5.2 μm together with an overlay of tracer trajectories. The liquid between the cells seems to be encapsulated in a vortex and to rotate as a torus with the axis of symmetry in x-direction. (b) Numerical simulation of a cluster. The streamlines indicate how tracers will be transported from the middle along a helical trajectories (see Fig.6). An experimental video from (a) can be found in supplementary material (Multimedia view)

Fig. 8 Period of cycles for a toroidal tracer motion between two RBCs as a function of the distance between the cells. Experimental data are represented by solid symbols and lines and numerical results by open symbols and dashed lines respectively. The fit represents a motion on an axial stretched torus with ellipses of revolution. It is based on equation (1) with $\pi/v_t$ as a fitting parameter and the minor axis is fixed to be $a = 3\mu m$ (see supplementary material).

Fig. 9 3D view showing the provenance (blue arrow) of the tracers and their trajectories for a cell velocity of 2.2 mm/s.

Fig. 10 Numerically calculated trajectories of a tracer escaping from flow vortices between two RBCs with velocity 2.2 mm/s where $X_0$ is the center between the two RBCs. (a) Temporal forward-backward motion along the direction of the flow. (b) Toroidal trajectory in the x-y plane, starting from the middle and going outwards.